\RequirePackage[right]{lineno}
\documentclass[aps,prl,preprint,nopacs,superscriptaddress]{revtex4}
\usepackage{graphicx}
\usepackage{physics}
\usepackage{braket}

\newcommand{\ene}{\varepsilon}

\begin{document}
\title{Probing quantum geometry through optical conductivity and magnetic circular dichroism}

\author{Barun Ghosh\footnote{Corresponding author (email): b.ghosh@northeastern.edu}}\affiliation{\footnotesize Department of Physics, Northeastern University, Boston, MA 02115, USA}

\author{Yugo Onishi}
\affiliation{\footnotesize Department of Physics, Massachusetts Institute of Technology, Cambridge, MA 02139, USA}

\author{Su-Yang Xu}\affiliation{\footnotesize Department of Chemistry and Chemical Biology, Harvard University, Massachusetts 02138, USA}

\author{Hsin Lin}
\affiliation{\footnotesize Institute of Physics, Academia Sinica, Taipei 11529, Taiwan}

\author{Liang Fu\footnote{Corresponding author (email): liangfu@mit.edu}}\affiliation{\footnotesize Department of Physics, Massachusetts Institute of Technology, Cambridge, MA 02139, USA}

\author{Arun Bansil\footnote{Corresponding author (email): ar.bansil@northeastern.edu}}
\affiliation{\footnotesize Department of Physics, Northeastern University, Boston, MA 02115, USA}

\begin{abstract}
Probing ground-state quantum geometry and topology through optical response is not only of fundamental interest, but it can also offer several practical advantages. Here, using first-principles calculations on antiferromagnetic topological insulator MnBi$_2$Te$_4$ thin films, we demonstrate how the generalized optical weight arising from the absorptive part of the optical conductivity can be used to probe the ground state quantum geometry and topology. We show that three septuple layers MnBi$_2$Te$_4$ exhibit an enhanced almost perfect magnetic circular dichroism for a narrow photon energy window in the infrared region. We calculate the quantum weight in a few septuple layers MnBi$_2$Te$_4$ and show that it far exceeds the lower bound provided by the Chern number. Our results suggest that the well-known optical methods are powerful tools for probing the ground state quantum geometry and topology.
\end{abstract}
\date{\today}
\maketitle

\vspace{0.5cm}
\textbf{Introduction}

In a crystalline solid, the quantum metric and the Berry curvature together constitute the complex quantum geometry of the Bloch wavefunction~\cite{Resta2011}. The quantum metric measures the gauge-invariant ``distance'' between Bloch wavefunctions at different momenta, while the Berry curvature characterizes the change in the phase of Bloch wavefunction along a closed contour in Brillouin zone. The quantum geometry of a solid can be directly manifested in its transport properties.
 The anomalous Hall conductivity can be expressed as an integral of the Berry curvature of the occupied bands over the Brillouin zone~\cite{tknn,vanderbilt_book}. Recently, it has been shown that Berry curvature dipole and quantum metric dipole lead to remarkable nonlinear transport phenomena~\cite{Fu_BCD,BCD_NLH,inhe_1,inhe_2,kaplan2023unification,quantum_metric_science,QM_longitudinal_nature} 

\begin{figure}[t]
\includegraphics[width=0.48\textwidth]{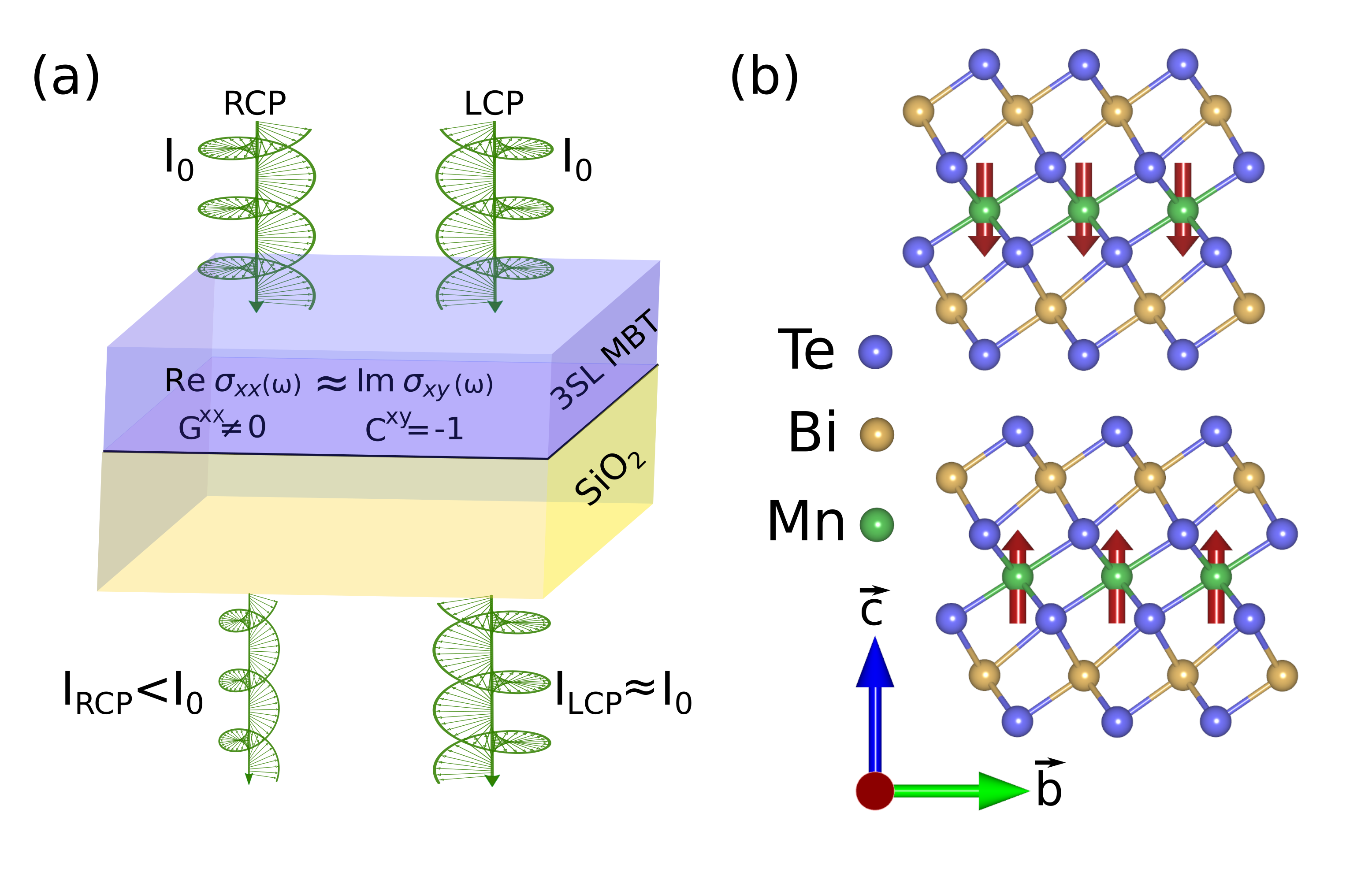}
\caption{(a) A schematic illustration of near-perfect magnetic circular dichroism (MCD) in 3SL  MnBi$_2$Te$_4$. The $I_0$ represents the equal intensity of right and left circularly polarized (RCP and LCP) light entering into the medium, while $I_{\rm RCP} (I_{\rm LCP})$ represents the intensity of the right (left) circularly polarized transmitted beam. When the perfect MCD occurs, the circularly polarized light only of a specific helicity is absorbed depending on the sign of ${\rm Im}~\sigma_{xy}$. (b) Crystal structure of two septuple layers (2SL) MnBi$_2$Te$_4$ in the AFM phase.}
\label{fig1}
\end{figure}

The effect of quantum geometry on the optical properties of solids has also received increasing attention~\cite{JM_shift_current,JM_CPGE,JA_PhysRevX.10.041041,Binghai_LMI,shg_QM_AA}. %, which are mainly determined by the photon-assisted interband transitions. 
The dipole moment matrix element for optical transition is closely related to the interband Berry connection~\cite{sipe_nonlinear}. Exploiting this link, recent studies have constructed an alternate description of optical nonlinear responses in the Riemannian geometry notations that involve both ground states and excited states~\citep{JA_natphys}. 

The connection between linear optical conductivity and {\it ground state quantum geometry} has remained largely unexplored. 
Recent theory \cite{Fu_fundamental_gap_prx} shows that the generalized optical weight, defined as the first negative moment of the absorptive part of optical conductivity (longitudinal and Hall) over the whole frequency range ($0<\omega<\infty$), is directly connected to the ground state quantum geometry and topology. This generalized optical weight is a complex quantity: its imaginary part, defined by magnetic circular dichroism (MCD), is connected to the ground state Chern number, while its real part defined by optical absorption, is connected to quantum metric through Souza-Willkins-Martin sum rule~\cite{souza_metric}. Specially, the real part of the generalized optical weight, recently termed the `quantum weight',  is directly determined by the$-$ quantum metric of the occupied band manifold integrated over the Brillouin zone. Although the quantum weight is an important ground state property that quantifies the degree of ``quantumness" of an insulating system, it has never been calculated for real materials to our knowledge. 

In this work,  %density functional theory (DFT) based 
we use first-principles calculations and effective field theory to study quantum geometry, optical absorption, and magnetic circular dichroism in two-dimensional MnBi$_2$Te$_4$, a magnetic topological insulator that exhibits trivial and Chern insulator ground states depending on the layer thickness~\cite{unique_mbt_prl,axion_chern_mbt}. We show by effective field theory that topological band inversion generally increases the quantum weight and therefore leads to a strong enhancement of infrared absorption. This is explicitly demonstrated by our first-principles calculation of the frequency-dependent optical conductivity and the generalized optical weight in MnBi$_2$Te$_4$. As the cutoff frequency increases, the generalized weight of MCD converges to the quantized Chern number, while the generalized weight of optical absorption converges to the quantum weight, which far exceeds the  lower bound provided by the Chern number \cite{Fu_fundamental_gap_prx}.
Finally, we show that the Chern insulator state in MnBi$_2$Te$_4$ exhibits an enhanced, almost perfect MCD for a narrow photon energy window in the infrared region.

\vspace{0.5cm}
\textbf{Optical conductivity and generalized optical weight}

For completeness, we first review recent results relating optical conductivity $\sigma_{\alpha\beta}(\omega)$ to quantum geometry and topology of the ground state \cite{Fu_fundamental_gap_prx}. 
We evaluate the two-dimensional linear optical conductivity using the Kubo-Greenwood formula for the non-interacting electronic systems as given by
\begin{equation}
\sigma_{\alpha\beta}(\omega)=\frac{e^2}{\hbar} \int{[d\bf k]} \sum_{n,m} f_{nm{\bf k}} \frac{i\epsilon_{mn{\bf k}}A^\alpha_{nm{\bf k}} A_{mn{\bf k}}^\beta}{\epsilon_{nm{\bf k}}+\hbar \omega+i\delta}.
\label{kubo}
\end{equation}
Here, $\epsilon_{n{\bf k}}$ is the energy eigenvalue of the $n^{th}$ Bloch state at crystal momenta ${\bf k}$; $[d{\bf k}]=d^2{\bf k}/(2\pi)^2$ in two dimensions. The interband Berry connection is given by $A^\alpha_{mn {\bf k}}=\braket{u_{m{\bf k}}|i \partial_\alpha|u_{n{\bf k}}}$, where $\ket{u_{n\bf{k}}}$ is the cell periodic part of the Bloch wavefunction. The difference in occupancy, $f_{nm\bf{k}}=f_{n\bf{k}}-f_{m\bf{k}}$, where $f_{n\bf{k}}$ is the Fermi distribution function for the $n^{th}$ band at $\bf{k}$, $\epsilon_{nm{\bf k}}=\epsilon_{n{\bf k}}-\epsilon_{m{\bf k}}$. The $\delta$ is an infinitesimal broadening parameter. Hereafter, we write the frequency-dependent optical conductivity $\sigma_{\alpha\beta}(\omega)$ as $\sigma_{\alpha\beta}$, omitting the argument $\omega$ for the sake of brevity.

The optical conductivity can be divided into the symmetric ($\sigma^L$) and the anti-symmetric ($\sigma^H$) parts: $\sigma^{L,H}$=($\sigma_{\alpha\beta} \pm \sigma_{\beta\alpha})/2$~\cite{powtw90,dichroric_fsum_vanderbilt}. The symmetric and anti-symmetric parts combined together form the absorptive (Hermitian) and the reactive (anti-Hermitian) parts:  $\sigma^{abs}=\mathrm{Re}~\sigma^L+ i \mathrm{Im} ~\sigma^H$, and $\sigma^{rea}=\mathrm{Re}~\sigma^H+ i \mathrm{Im}~\sigma^L$.  In this work, our primary focus is on $\sigma^{abs}$; its real part ( $\mathrm{Re}~\sigma^L$ ) is related to the absorption of linearly polarized light, while its imaginary part ($\mathrm{Im}~\sigma^H$) is responsible for magnetic circular dichroism.

We now explore the connection between optical conductivity and ground-state quantum geometry. The quantum geometric tensor for a set of $N$ bands parametrized by the wavevector ${\bf k}$ can be expressed as
\begin{equation}
\mathcal{Q}_{\alpha\beta}^{{ij{\bf k}}}=\braket{\partial_\alpha u_{i\bf k}|(1-P_{\bf k})|\partial_\beta u_{j\bf k}}.
\end{equation}
Here, $i,j=1,....,N$. The projection operator $P_{\bf k}$ is given by $P_{\bf k}=\sum_{i=1}^N \ket{u_{i\bf k}}\bra{u_{i\bf k}}$. The real and the imaginary part of $\mathcal{Q}^{\alpha\beta}_{ij}$  represent the non-Abelian quantum metric ($G^{\alpha\beta}$) and the non-Abelian Berry curvature ($F^{\alpha\beta}$), i.e. $\mathcal{Q}^{\alpha\beta}=G^{\alpha\beta}-\frac{i}{2}F^{\alpha\beta}$. %~\cite{Abelian_NA_QGT}.
Following Ref~\citep{Fu_fundamental_gap_prx}, we define the generalized optical weight using the first negative moment of the absorptive part of the optical conductivity as
\begin{equation}
W^1_{\alpha\beta}(\omega_c)=\int_0^{\omega_c} d\omega \frac{\sigma_{\alpha\beta}^{abs}}{\omega},
\label{weight}
\end{equation}
here, $\omega_c$ is a cutoff frequency; in the $\omega_c \rightarrow \infty$ limit, $W^1_{\alpha\beta}$ represents the full spectral weight. For an insulator, $\sigma_{\alpha\beta}^{abs}$ is non-zero only for photon energy higher than the band gap; thus, the above integral is convergent. The $\sigma_{\alpha\beta}^{abs}$ can be recovered from $W^1(\omega_c)$ using the formula $\sigma_{\alpha\beta}^{abs}(\omega)=\omega dW^1_{\alpha\beta}/d\omega$, and $\sigma_{\alpha\beta}^{rea}$ can be obtained from $\sigma_{\alpha\beta}^{abs}$ using the Kramers-Kronig relations. Thus, $W^1(\omega_c)$ represents an important quantity that carries all the essential information about the optical conductivity; it represents the integrated dielectric loss below the photon energy $\hbar\omega_c$.

To establish a direct connection between the $\sigma_{\alpha\beta}^{abs}$ and the quantum geometry we express
\begin{equation}
\sigma_{\alpha\beta}^{abs}(\omega)=\pi\omega e^2 \int[d{\bf k}] \sum_{n,m} f_{nm{\bf k}}A^\alpha_{nm{\bf k}} A^\beta_{mn{\bf k}}\delta(\epsilon_{mn{\bf k}}-\hbar\omega).
\label{sigma_abs}
\end{equation}
Combining Eq.~\ref{weight} and Eq.~\ref{sigma_abs}, and using $f_o=1$ for the occupied (o) states and $f_u=0$ for the unoccupied (u) state at zero temperature, the $W^1_{\alpha\beta}(\omega_c)$ can be written as an integral in the ${\bf k}-$space
\begin{equation}
\begin{split}
W_{\alpha\beta}^1(\omega_c)=\int_0^{\omega_c} d\omega \frac{\sigma_{\alpha\beta}^{abs}}{\omega}\\
=\frac{\pi e^2}{\hbar} \sum_{o,u}\int_{\epsilon_{uo}\leq\hbar\omega_c} [d{\bf k}] A_{ou}^\alpha A_{uo}^\beta.
\end{split}
\label{W1kspace}
\end{equation}
For a small cutoff frequency $\omega_c$, only a limited region of the ${\bf k}$-space and a subset of the bands that satisfies the condition $\epsilon_{uo}\leq \hbar\omega_c$ contributes to the BZ integral. In the $\omega_c \rightarrow \infty$ limit, the frequency integral represents the full spectral weight when the optical transition involves all the bands, and the ${\bf k}$-space integral encompasses the entire BZ. 

In this work, we focus on a material that respects the $\mathcal{C}_{3z}$ rotation symmetry, which ensures $\sigma^L=\sigma_{xx}=\sigma_{yy}$ and $\sigma^H=\sigma_{xy}=-\sigma_{yx}$, and it is assumed for the rest of our discussion. Due to this, hereafter, we drop the superscript $L$, $H$ and write the symmetric and the anti-symmetric parts of the optical conductivity as $\sigma_{xx}$ and $\sigma_{xy}$, respectively. The relevant real and imaginary part of $W_{\alpha\beta}^1$  leads to

\begin{equation}
{\rm Re}~ W^1_{xx}(\omega_c\rightarrow \infty)\equiv\int_0^{\omega_c \rightarrow \infty} d\omega \frac{{\rm Re} ~\sigma_{xx}}{\omega}= \frac{e^2}{2\hbar} K_{xx},
\label{EqnsigmaL}
\end{equation}

\begin{equation}
{\rm Im} ~W^1_{xy}(\omega_c\rightarrow \infty) \equiv \int_0^{\omega_c \rightarrow \infty} d\omega \frac{{\rm Im}~ \sigma_{xy}}{\omega}=-\frac{e^2}{4\hbar} C_{xy}.
\label{EqnsigmaH}
\end{equation}
Here, $K_{xx}$ is the quantum weight, a quantum property of the insulating ground state, given by $K_{xx}=2\pi\int[d{\bf k}]g_{xx}$; $g_{xx}$= $\mathrm{Tr}[G_{xx}]$ is the trace of the non-Abelian quantum metric of the occupied band manifold, or alternatively, it represents the Abelian quantum metric of the Slater determinant states formed by the occupied bands.  The quantum weight is related to electron localization length in an insulator \cite{souza_metric}, and it represents a quantitative measure of the degree of ``quantumness" in the insulating state~\citep{Fu_fundamental_gap_prx}. The Chern number ($C_{xy}$) in two dimensions is given by the BZ integral of the Berry curvature of the occupied band manifold: $C_{xy}=2\pi \int[d{\bf k}]\sum_oF_{xy}^o$.  

The quantum metric and the Berry curvature of the Slater determinant state of the occupied band manifold must satisfy the following inequality \cite{roy, torma}
\begin{equation}
g_{xx}+g_{yy} \geq \abs{F_{xy}}.
\label{QMBC_ineq}
\end{equation}
This leads to a lower bound on the quantum weight in a two-dimensional system
\begin{equation}
K \equiv K_{xx}+K_{yy} \geq \abs{C_{xy}}.
\label{KC_ineq}
\end{equation}

We note that, in the special case of a single occupied band, the quantum geometric tensor is Abelian. Then, if, the so-called ideal metric condition $\Tr~g=\abs{F_{xy}}$ is satisfied at every ${\bf k}$, the quantum weight equates the lower bound provided by the Chern number: $K=\abs{C}$~\cite{PhysRevLett.114.236802,PhysRevLett.127.246403,ledwith2022vortexability}.
Armed with these notations, we now study the optical weights and their connection to ground-state quantum geometry and discuss how they lead to enhanced infrared absorption and a near-perfect MCD in a real material.  

\vspace{1.5cm}
\textbf{Crystal structure and electronic structure of MnBi$_2$Te$_4$}
\begin{figure}[t]
\includegraphics[width=0.7\textwidth]{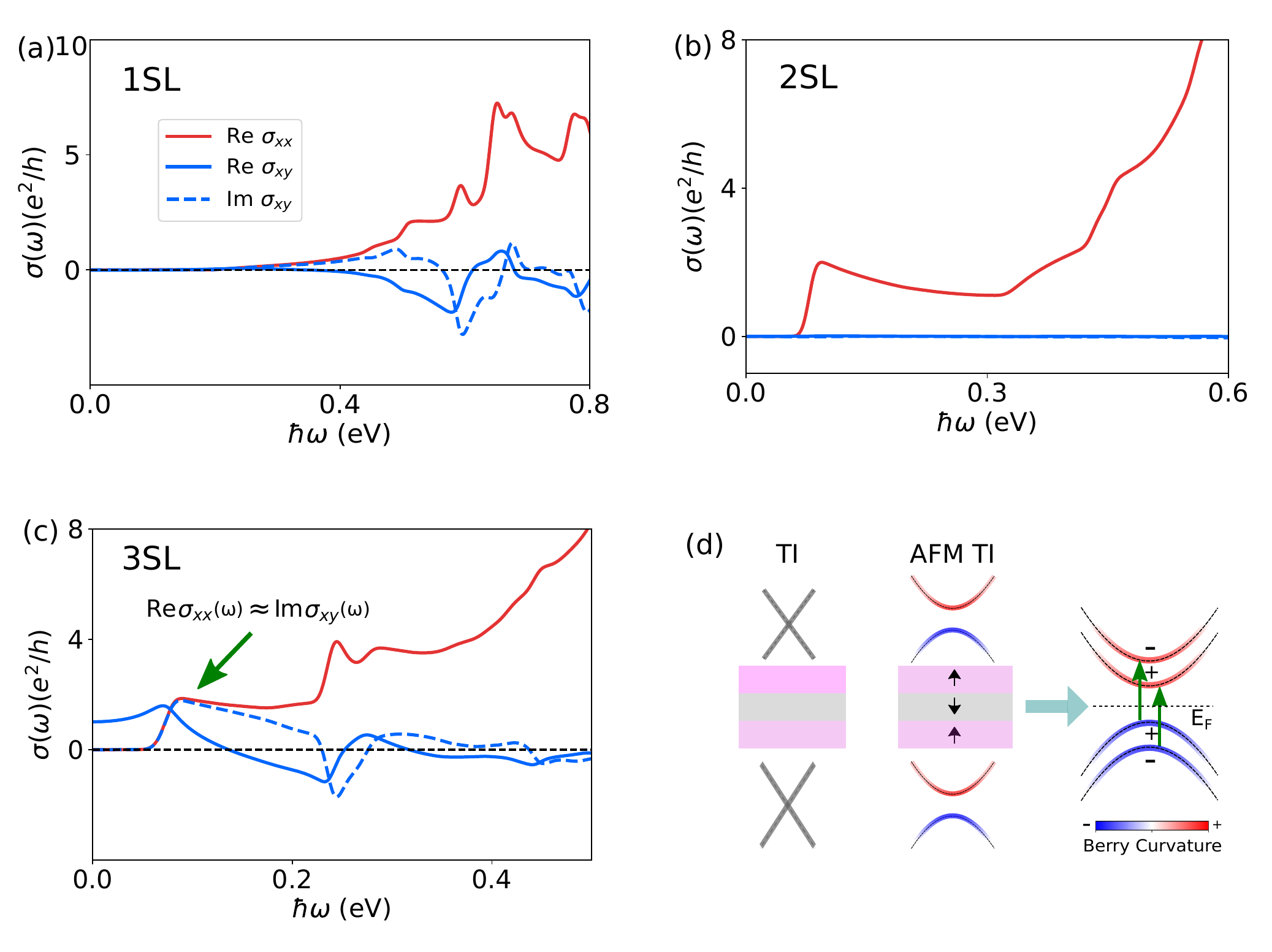}
\caption{ {\bf Optical conductivity of a few SL thick MnBi$_2$Te$_4$}. (a) 1SL, (b) 2SL, and (c) 3SL MnBi$_2$Te$_4$. For 1SL, the ${\rm Re}~\sigma_{xx}$ increases gradually, while for 2SL and 3SL it has a sudden onset. In 2SL $\mathcal{PT}$ symmetry forbids $\sigma_{xy}$ at all frequencies. (d) A schematic illustration of the low energy optical transitions involving the gapped Dirac cone states of 3SL MnBi$_2$Te$_4$. The band splitting arises from the uncompensated magnetization and the hybridization between the top and the bottom Dirac cone states. The $\pm$ sign denotes the parity of the bands. A direct optical transition involving the highest valence band and the lowest conduction band is strongly suppressed due to the optical selection rule.}
\label{fig2}
\end{figure}

Our study is focused on MnBi$_2$Te$_4$ which recently emerged as the first stoichiometric compound to host an antiferromagnetic topological insulator state~\cite{Fu_TI_Z2,JM_afm_ti,bahadur_review,Otrokov2019,mbt1_theory,jing_wang_model_mbt,layerhall,PhysRevLett.122.206401,mbt_review}. In the non-magnetic phase, it crystallizes in the $R\bar{3}m$ space group. It has a layered crystal structure, where individual MnBi$_2$Te$_4$ septuple layer (SL) building blocks, in a sequence of -Te-Bi-Te-Mn-Te-Bi-Te- atomic layers, are vertically stacked and stabilized via weak van der Wall's attraction force. Below the magnetic transition temperature, the Mn spins favor an in-plane ferromagnetic coupling (see Fig.~\ref{fig1}(b)) and a Neel-type antiferromagnetic ordering in the vertical direction~\cite{mbt_crystal_str}. 
The thin films with an odd number of SL preserve the inversion symmetry ($\mathcal{P}$), while the even number of SL breaks the inversion and time-reversal symmetry ($\mathcal{T}$) but it preserves the combined $\mathcal{PT}$ symmetry. Additionally, the $\mathcal{C}_{3z}$ rotational symmetry is always present, irrespective of the number of SL.

The low energy band structure of a few SL MnBi$_2$Te$_4$ has features related to the surface states of a topological insulator thin film, where two gapless Dirac cones reside on opposite surfaces ~\cite{model_ham_ti}. In a few SL MnBi$_2$Te$_4$ (see Fig.~\ref{fig2}(d)), the band gap at the Dirac crossing arises from two factors: through hybridization between the top and the bottom surface Dirac cone states and due to the exchange coupling of the Mn spins.  The uncompensated magnetization in odd SL results in singly degenerate spin-split bands, while in the even SL, $\mathcal{PT}$ symmetry ensures the double degeneracy of the bands at every crystal momenta. The 1SL MnBi$_2$Te$_4$ is topologically trivial, while in 3SL, the gapped Dirac cone states from the top and the bottom surface each contribute $e^2/2h$ to the Hall conductivity \cite{Fu_TI_Z2}, leading to a quantum anomalous Hall insulator phase with a Chern number $|C|=1$ ~\cite{unique_mbt_prl,mbt_qah,mbt_cri3}. The even SL MnBi$_2$Te$_4$ hosts the so-called zero Hall plateau state~\cite{unique_mbt_prl,layerhall,axion_chern_mbt}. 

We now proceed to compute the optical conductivity and the optical weights for a few SL MnBi$_2$Te$_4$ within the Wannier function-based tight-binding framework (see SM for details)~\cite{wannier_rmp}. The relevant Hilbert space is spanned by bands of Bi $p$, Te $p$, and Mn $d$ orbital characters. The optical response thus arises due to the interband transitions between these finite numbers of occupied and the unoccupied band manifold. First, we focus on the low-frequency response and then discuss the high-frequency behavior of the optical conductivity and the optical weights. 

%\subsection{Low frequency response}
\vspace{0.5cm}
\textbf{Low frequency response}

The optical conductivity of a few SL thick MnBi$_2$Te$_4$ films for photon energy in the infrared region is shown in Fig.~\ref{fig2}(a)-(c). The $\sigma^{abs}$ is directly connected to the interband transitions; thus ${\rm Re}~\sigma_{xx}$ and ${\rm Im}~\sigma_{xy}$ both vanish identically for photon energy lower than the band gap. For 1SL MnBi$_2$Te$_4$, the ${\rm Re}~\sigma_{xx}$ and ${\rm Im}~\sigma_{xy}$ increase gradually with frequency, while for 2SL and 3SL, they have a sudden onset. The ${\rm Re}~\sigma_{xx}$ displays universal characteristics of optical conductivity in a two-dimensional system, where an interband transition peak is followed by a plateaulike region. Because of the broken $\mathcal{T}$ symmetry, 1SL and 3SL have a finite $~\sigma_{xy}$, while in 2SL,  the combined $\mathcal{PT}$ symmetry forbids $\sigma_{xy}$ at all frequencies. The 3SL MnBi$_2$Te$_4$ hosts the quantum anomalous Hall insulator phase, and in the $\hbar\omega \rightarrow 0$ limit, ${\rm Re}~\sigma_{xy}$ starts from the quantized value of $e^2/h$ and develops a prominent peak. As indicated by the green arrow in Fig.~\ref{fig1}(c),  the ${\rm Re}~\sigma_{xx}\approx {\rm Im}~\sigma_{xy}$ in a narrow frequency window  in 3SL MnBi$_2$Te$_4$. This has important consequences on the MCD, which we will discuss later. 

In 3SL, the uncompensated magnetization and the hybridization between the top and the bottom surface Dirac cones lead to singly degenerate spin split bands (see Fig.~\ref{fig2}(d)). The low-frequency optical response of a few SL MnBi$_2$Te$_4$ mostly arises from the interband transitions involving these gapped Dirac cone surface states. We note that a direct optical transition between the highest valence band and the lowest conduction band is strongly suppressed due to the optical selection rule. Since the system is close to the topological phase transition, and the low-frequency response arises from the gapped Dirac surface states, we study the optical conductivity of a gapped Dirac cone model in the next section.

\begin{figure*}[t]
\includegraphics[width=0.95\textwidth]{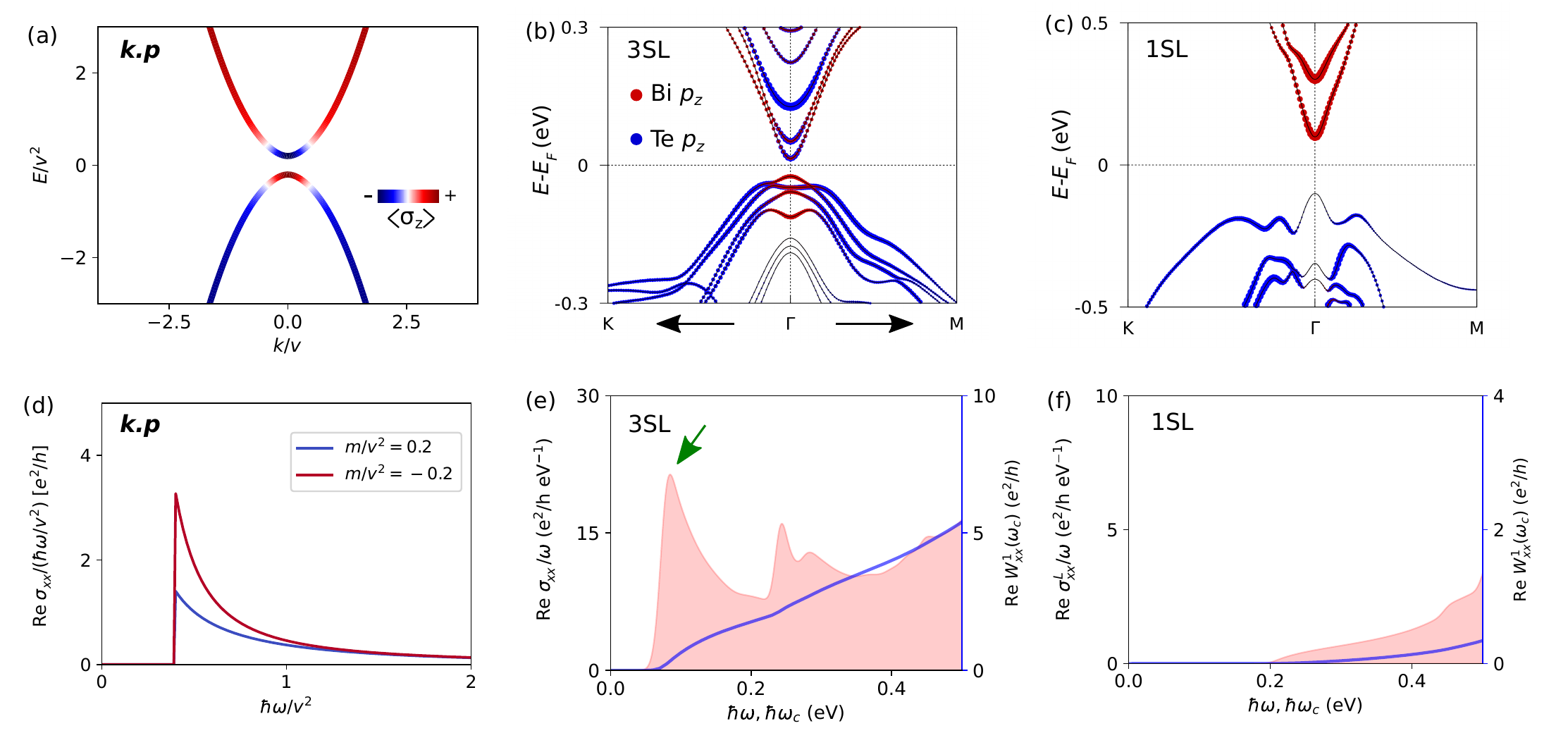}
\caption{{\bf Enhanced infrared absorption due to topological band inversion.} (a) Band inversion in the ${\bf k.p}$ model. Color represents the spin polarization of the bands. (b,c) Orbital resolved band structure of (b) 3SL, (c) 1SL MnBi$_2$Te$_4$. The 3SL MnBi$_2$Te$_4$ hosts band inversion between the Bi $p_z$ and Te $p_z$ orbitals near the $\Gamma$ point, which is absent in 1SL MnBi$_2$Te$_4$. (d) The ${\rm Re}~\sigma_{xx}/\omega$ calculated from the ${\bf k.p}$ model in the topological and non-topological phase. (e,f) The ${\rm Re}~\sigma_{xx}/\omega$ and the corresponding optical weight ${\rm Re} ~W^1_{xx}$ for (e) 3SL and (f) 1SL MnBi$_2$Te$_4$. In our ${\bf k.p}$ model, there is a clear enhancement of  ${\rm Re}~\sigma_{xx}/\omega$ in the topological phase ($m<0$) compared to the trivial phase ($m>0$) for the same magnitude of the gap ($2|m|$). This is further supported by our first principles results: for 1SL MnBi$_2$Te$_4$ ${\rm Re}~\sigma_{xx}/\omega$ increases gradually, while for the 3SL MnBi$_2$Te$_4$, it has a sharp peak in the low-frequency region.}
\label{figEIA}
\end{figure*}

\vspace{0.5cm}
\textbf{Enhanced optical absorption due to band inversion}

In this section, we study the gapped Dirac cone model as the low-energy effective theory of systems near topological phase transition. We start with the following Dirac Hamiltonian,
\begin{align}
    H &= (-m-k^2)\sigma_z + v(k_x\sigma_x + k_y\sigma_y),
\end{align}
where $m$ is the mass and $v$ is the velocity of the Dirac fermion. In this model, topological phase transition can be controlled by tuning the mass parameter $m$; the Dirac Hamiltonian describes a topologically trivial (nontrivial) phase when $m>0 (m<0)$. The optical absorption of linearly polarized light is described by the real part of the longitudinal conductivity, ${\rm Re}~\sigma_{xx}$.  We obtain an exact analytical expression of the ${\rm Re}~\sigma_{xx}$ for our Dirac model
\begin{align}
    {\rm Re}~\sigma_{xx}(\omega) &= \frac{\omega e^2}{16v^2} \int_0^\infty k\dd{k} \delta(\ene_+(\vec{k})-\ene_-(\vec{k})-\hbar\omega) \nonumber \\
    &\quad \times\frac{2M^2 + 2(k/v)^4+(k/v)^2}{(M+(k/v)^2)^2 + (k/v)^2} \nonumber \\
    &= \frac{e^2}{4\hbar \Omega^2} \sum_{i}\frac{2M^2 + 2q_i^4 + q_i^2}{\abs{2M + 2q_i^2 + 1}} \label{eq:Dirac},
\end{align}
where $M=m/v^2, \Omega=\hbar\omega/v^2$ are the renormalized mass and frequency, and $\ene_{\pm}=\pm\sqrt{(m+k^2)^2 + v^2 k^2}$ is the energy dispersion of the Dirac fermion. The $q_i=k_i/v$ is a renormalized wavevector at which the resonance occurs and is a function of the frequency. The explicit form of $q_i$ is provided in the SM. 

As shown in Fig.~\ref{figEIA}(d), the topological phase ($m < 0$) clearly exhibits a larger optical response than the trivial phase for the same value of the gap ($2\abs{m}$). This can be seen more explicitly from the Eq.~\eqref{eq:Dirac}. For simplicity, let us consider the response at the band edge when $m\le v^2/2$. At frequency $\omega_0=2m/\hbar$, the optical transition occurs at $k=0$,  and hence ${\rm Re}~\sigma_{xx}(\omega_0) \propto \abs{v^2+2m}^{-1}$. Therefore, the optical conductivity at the band edge is always larger for $m<0$ than for $m>0$ for the same value of $\abs{m}$. This indicates that the band inversion enhances the optical response near the topological phase transition.
% In particular, the optical response diverges as the $m=v^2/2$ limit is approached when the band dispersion becomes flat around $k=0$. 

This is well supported by our first-principles calculations on a few SL MnBi$_2$Te$_4$.  As shown in Fig.~\ref{figEIA}(b-c), the 3SL MnBi$_2$Te$_4$ host a clear band inversion between the Bi $p_z$  and Te $p_z$ orbitals near the $\Gamma$ point, while the band inversion is absent in 1SL due to the quantum confinement effect. In 1SL MnBi$_2$Te$_4$, the ${\rm Re}~\sigma_{xx}/{\omega}$  increases gradually with frequency, while for 3SL, it shows a sudden onset in the low-frequency region, indicating an enhanced infrared absorption and a larger quantum weight. 

The band inversion-induced enhancement in optical absorption can be understood from the presence of a large quantum metric carried by the low energy bands of 3SL MnBi$_2$Te$_4$. In 3SL, because of the band inversion, the wavefunction of the low energy bands near the $\Gamma$ point changes rapidly between two nearby ${\bf k}$ points, resulting in a large quantum metric in this region. The ${\rm Re}~\sigma_{xx}/{\omega}$ is directly connected to the quantum metric, and therefore, band inversion generally leads to enhanced optical absorption.

%This enhancement in optical absorption in the topological phase can be further formalized following the inequality between the quantum weight and the Chern number. The quantum weight arises from the weight of optical absorption, while the Chern number is the weight of the MCD. As per Eq.~\ref{KC_ineq}, the Chern number $(C)$ is the lower bound of the quantum weight $(K)$: $K\ge\abs{C}$. Therefore, the topologically nontrivial state with a non-zero Chern number must have a larger lower bound of the quantum weight, resulting in an enhanced optical absorption in the low-frequency region. 

\vspace{0.5 cm}
\textbf{Chern number as the weight of MCD\label{sec_mcd}}

\begin{figure}[t]
\includegraphics[width=0.75\textwidth]{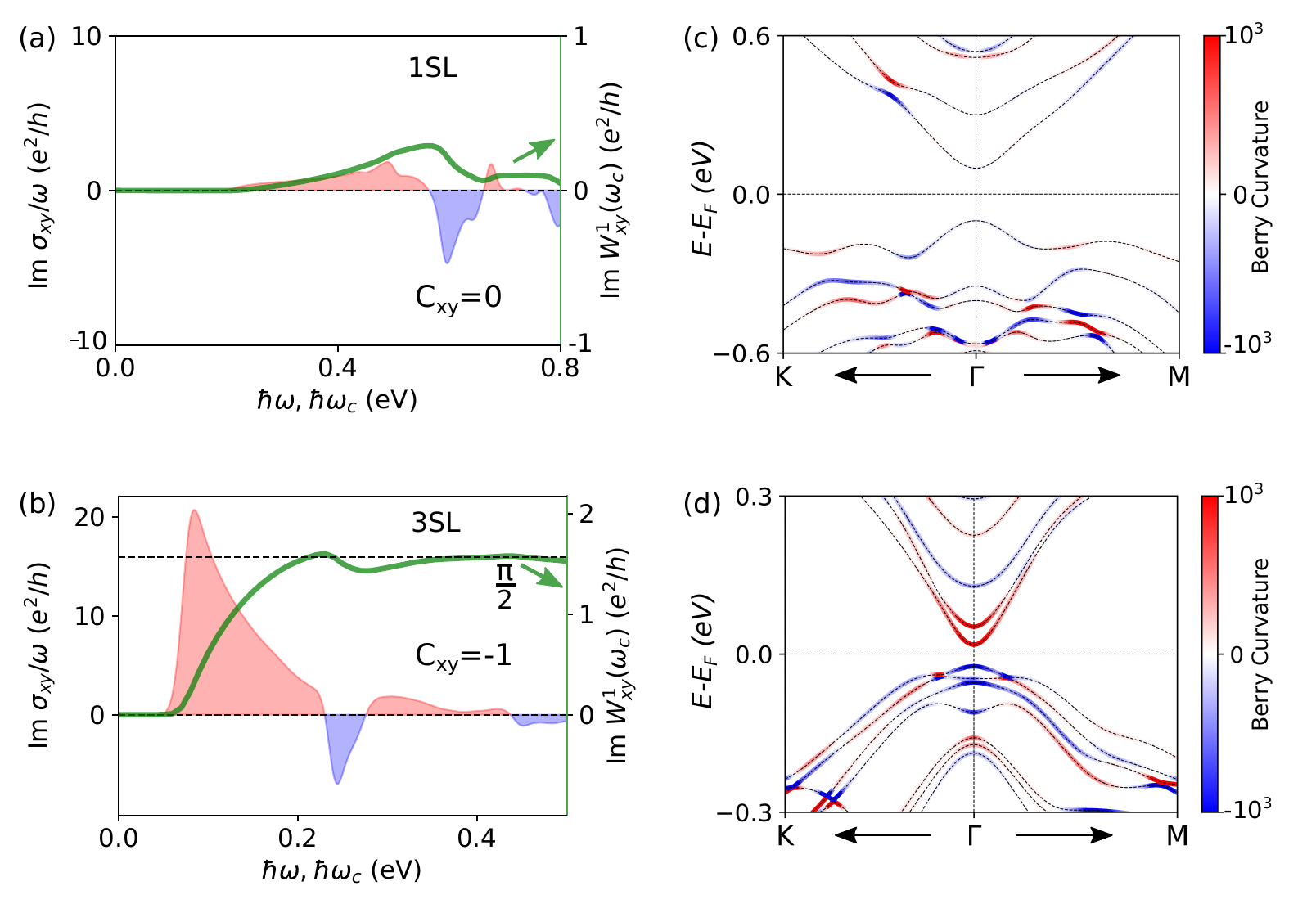}
\caption{{\bf Imaginary part of the optical weight and its connection to the Chern number.} (a,b) The ${\rm Im}~\sigma_{xy}/\omega$ $({\rm Im}~W^1_{xy})$ varying photon energy (cutoff energy) for (a) 1SL and (b) 3SL MnBi$_2$Te$_4$. In the case of 1SL, the ${\rm Im}~W^1_{xy}$ converges to zero, while for the 3SL, the ${\rm Im}~W^1_{xy}$ converges to $e^2/4\hbar$, revealing their trivial and Chern insulator ground state, respectively. (c,d) The band and momentum resolved Berry curvature distribution for (c) 1SL and (d) 3SL MnBi$_2$Te$_4$. In 3SL MnBi$_2$Te$_4$, the low energy bands carry large Berry curvature around a small region near the $\Gamma$ point.}
\label{fig4}
\end{figure}

We now focus on the optical weight arising from the ${\rm Im}~\sigma_{xy}$, which is responsible for the MCD. As shown in Fig.~\ref{fig4}(a)-(b), for 1SL, the ${\rm Im}~\sigma_{xy}/\omega$ increases gradually with the photon energy and shows an oscillatory behavior. In contrast, for 3SL, the ${\rm Im}~\sigma_{xy}/\omega$ has a large peak at $\hbar\omega \approx 85$ meV, and it displays a diminishing oscillatory pattern with increasing frequency. The corresponding optical weight, ${\rm Im}~W^1_{xy}$, for 1SL MnBi$_2$Te$_4$, starts from zero and gradually increases with the cutoff frequency before approaching the horizontal axis, indicating the trivial nature of 1SL. In contrast, for 3SL, ${\rm Im}~W^1_{xy}$ starts from zero and it quickly approaches the quantized value of $e^2/4\hbar$, revealing the Chern insulator ground state ($C_{xy}=-1$) of 3SL MnBi$_2$Te$_4$. Interestingly, the weight arising from the first peak of ${\rm Im}~\sigma_{xy}/\omega$ is sufficient to saturate the quantized Chern number of 3SL MnBi$_2$Te$_4$. 

This rapid convergence of ${\rm Im}~W^1_{xy}$ with the cutoff frequency for 3SL MnBi$_2$Te$_4$ can be understood from the Berry curvature distribution in ${\bf k}$-space. As described above, the low energy bands of 3SL MnBi$_2$Te$_4$ are the gapped Dirac cone surface state of a magnetic topological insulator thin film, and they carry a large Berry curvature that is highly concentrated in a small ${\bf k}$-space region (see Fig.~\ref{fig4}(d)) near the $\Gamma$ point. The optical transition at low frequency mostly involves these gapped Dirac-like bands. Therefore, following the Eq.~\ref{W1kspace}, for a cutoff energy of $\hbar\omega_c \approx 23$ meV, the low energy bands around a small ${\bf k}$-space region near the $\Gamma$ point that satisfies $\hbar\omega_c \leq \epsilon_{uo\bf k}$ is sufficient to saturate the quantized Chern number of 3SL MnBi$_2$Te$_4$.

\vspace{0.5cm}
\textbf{Optical rotations and perfect MCD}

Next, we quantify the MCD arising from ${\rm Im}~\sigma_{xy}$ and show that 3SL MnBi$_2$Te$_4$ in the Chern insulator phase exhibit an enhanced near-perfect MCD for a narrow photon energy window in the infrared region. For completeness, we also quantify the complex Faraday and Kerr rotation angles that arise from a non-zero $\sigma_{xy}$. The real part, $\theta$, of the complex Kerr (Faraday) angle is representative of the rotation in the plane of linearly polarized incident light after reflection (transmission), and it is directly related to the ${\rm Re}~\sigma_{xy}$; the imaginary part, $\eta$, represents the ellipticity of the reflected (transmitted) beam, and it arises from the ${\rm Im}~\sigma_{xy}$, which in turn lead to circular dichroism. Using the frequency-dependent rotation angles and the MCD, the $\sigma_{xy}$ and $\sigma_{xx}$ can be obtained. 

\begin{figure}[t]
\includegraphics[width=0.8\textwidth]{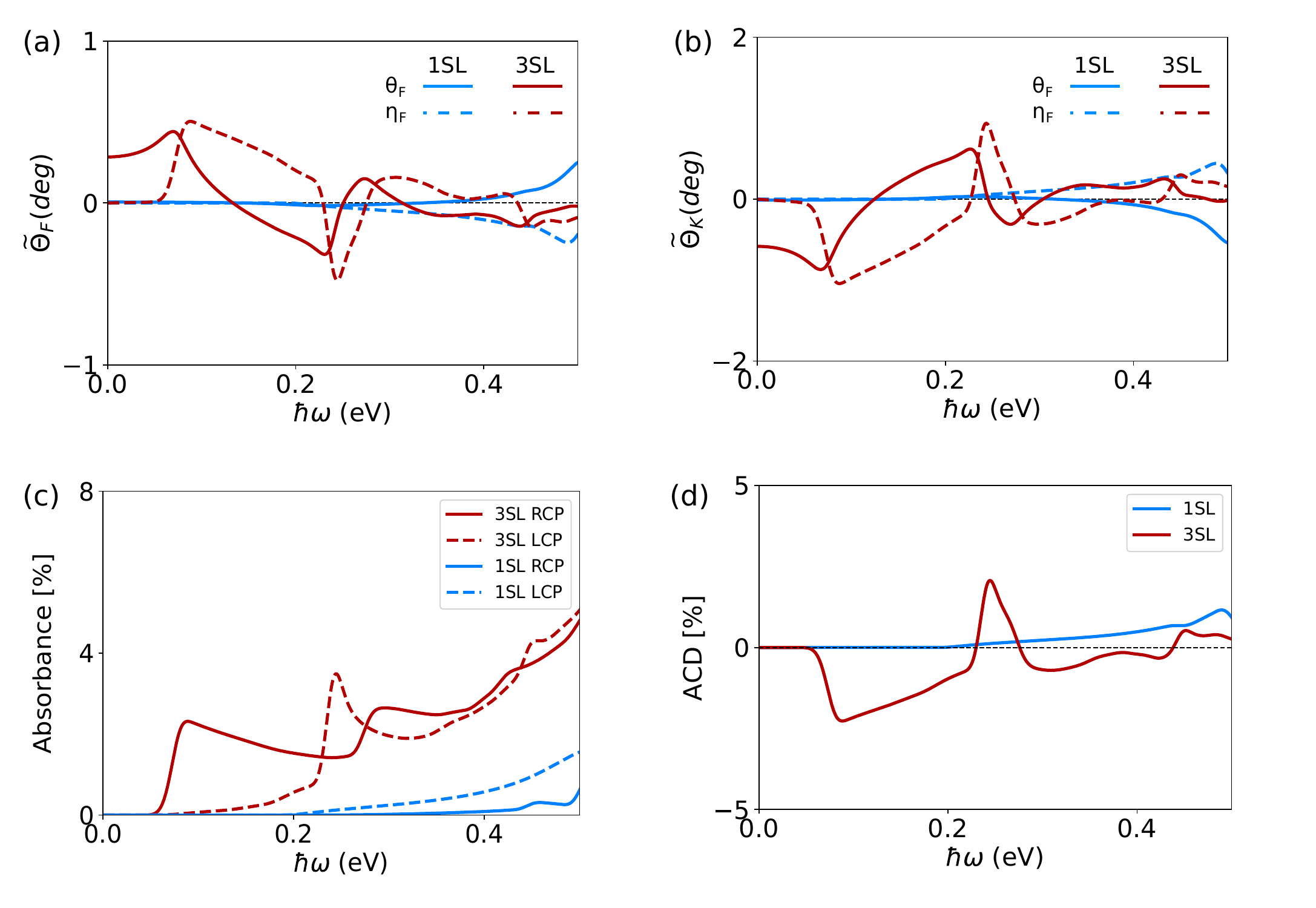}
\caption{{\bf Optical rotations and near perfect magnetic circular dichroism.} The complex (a) Faraday angles and (b) Kerr angles for 1SL and 3SL MnBi$_2$Te$_4$. (c) The absorption probability of circularly polarized light of different helicity for 1SL and 3SL MnBi$_2$Te$_4$. The 3SL MnBi$_2$Te$_4$ exhibits almost perfect MCD for $65 \lesssim \hbar\omega \lesssim 150$ meV. (d) The resultant absorptive CD.}
\label{fig5}
\end{figure}

We consider a realistic experimental setup, where the MnBi$_2$Te$_4$ thin film on a SiO$_2$ substrate is placed in a vacuum (see Fig.~\ref{fig1}(a)). For the normal incidence of linearly polarized light, the complex Faraday ($\tilde{\Theta}_F$) and Kerr ($\tilde{\Theta}_K$) angles in the thin film limit can be approximated as~\cite{kerr_0,kerr_1,haldane_cd,perfect_cd_haldane}
\begin{equation}
\tilde{\Theta}_F=\theta_F+i\eta_F=\frac{\mu_0c\sigma_{xy}}{1+n_{sub}+\mu_0c\sigma_{xx}},
\label{faraday}
\end{equation}
\begin{equation}
\tilde{\Theta}_K=\theta_K+i\eta_K=\frac{2\mu_0c\sigma_{xy}}{1-(n_{sub}+\mu_0c\sigma_{xx})^2}.
\label{kerr}
\end{equation}
Here, $\sigma_{xy}$ and $\sigma_{xx}$ are the two-dimensional conductivities in $\Omega^{-1}$ unit. The $n_{sub}$  is the refractive index of the substrate, and the incident medium is the vacuum.  We use  $n_{sub}=\sqrt{\epsilon^{SiO_2}_{xx}}=1.97$ in our calculations.  In the limiting case, $\theta_F \propto {\rm Re} ~\sigma_{xy}$ and $\eta_F \propto {\rm Im}~\sigma_{xy}$. 
As shown in Fig.~\ref{fig5}(a), the photon energy dependence of $\theta_F$ and $\eta_F$ largely follows the ${\rm Re}~\sigma_{xy}$ and ${\rm Im}~\sigma_{xy}$, respectively. In the $\hbar\omega \rightarrow 0$ limit, for 1SL, the $\theta_F$ vanishes  due to its trivial nature, while for the 3SL, the $\theta_F$ has a finite value ($\approx 0.28^\circ$). This value depends on the substrate refractive index, and in the case of a freestanding 3SL MnBi$_2$Te$_4$, in $\hbar\omega \rightarrow 0$ limit, $\theta_F$ attains the universal quantized value of $\theta_F=(\alpha_{fine})\approx 0.42^\circ$; here $\alpha_{fine}(\approx1/137)$ is the fine-structure constant~\cite{quantized_kerr_farady_NPA,topological_quantization_scz,quantized_faraday_mcdonald,kerr_faraday_mbt}. The complex Kerr angle shares a similar pattern as the Faraday angle, although it has an opposite relative sign. Depending on the substrate refractive index and the choice of the broadening parameter ($\delta$), the $\theta_K$ can reach values $\sim 1^\circ$, comparable to other magnetic 2D materials~\cite{Yang2020}. As evident from the $\eta_K$ and $\eta_F$ results, the transmitted and the reflected light both become elliptically polarized. The larger magnitude of $\eta_K$ in comparison to $\eta_F$ suggests that the reflected light attains a higher degree of ellipticity than the transmitted beam. The non-zero $\eta$ results in magnetic circular dichroism. 

We now focus on the MCD. The absorption probability for circularly polarized light of helicity $\hat{\chi}$ ($\hat{\chi}=+1$ for LCP, and $\hat{\chi}=-1$ for RCP) for MnBi$_2$Te$_4$ thin film on a substrate can be calculated using
\begin{equation}
A_{\hat{\chi}}=\frac{4\mu_0c({\rm Re}~\sigma_{xx}-\hat{\chi}{\rm Im}~\sigma_{xy})}{|1+n_{sub}+\mu_0c(\sigma_{xx}+i\hat{\chi}\sigma_{xy})|^2} .
\end{equation}
As evident from Fig.~\ref{fig5}(c), circularly polarized lights of opposite helicity are absorbed differently in few SL MnBi$_2$Te$_4$ due to the presence of a finite $\sigma_{xy}$. The resultant absorptive CD ($ACD=A_{LCP}-A_{RCP}$) is quantified in Fig.~\ref{fig5}(d). In 1SL, we observe a gradual increase in the absorbance for photon energy higher than the band gap. In contrast, an almost perfect MCD is observed for 3SL MnBi$_2$Te$_4$ in the photon energy range $65\lesssim \hbar\omega \lesssim 150$ meV. We note that, in this photon energy window, the ${\rm Re}~\sigma_{xx}$ is nearly equal to the ${\rm Im}~\sigma_{xy}$, resulting in the absorption of only RCP light, while the absorption probability for the LCP light almost vanishes. Although at $\hbar\omega \approx 85$ meV, the absorption probability for RCP reaches the maximum value, it is still only $\sim$ 2.3\% due to the two-dimensional nature of the few-layered MnBi$_2$Te$_4$. This value can be enhanced by adjusting the refractive indices of the incident medium and the substrate. The 3SL MnBi$_2$Te$_4$ with the opposite magnetic configuration ($\uparrow\downarrow\uparrow$ vs. $\downarrow\uparrow\downarrow$) only absorbs LCP light instead of RCP light. The LCP and RCP lights are absorbed almost equally for higher photon energy, resulting in a small CD. 

This perfect MCD in 3SL MnBi$_2$Te$_4$ can be understood from the absence of dissipation through Joule heating in a two-dimensional material for incident light of a specific helicity. For circularly polarized light of helicity $\hat{\chi}$, the power dissipation through Joule heating in a 2D conductor can be estimated as
\begin{equation}
P_{\hat{\chi}}={\rm Re}({\bf j^*.E}) \propto {\rm Re}~(\sigma_{xx}+\sigma_{yy})-2\hat{\chi}{\rm Im}~\sigma_{xy}. 
\end{equation}
Here, ${\bf j}$ is the surface current density induced by the electric field ${\bf E}$ of the circularly polarized incident light. Clearly, in $\mathcal{C}_{3z}$ symmetric 3SL MnBi$_2$Te$_4$, the $P_{\hat{\chi}}$ vanishes for circularly polarized light of helicity $\hat{\chi}$ when ${\rm Re}~\sigma_{xx}=\hat{\chi}{\rm Im}~\sigma_{xy}$, resulting in the absorption of circularly polarized light of opposite helicity $-\hat{\chi}$. In terms of the quantum geometry, for a two-band system with a single occupied band, this requires the equality $g_{xx}({\bf k})=\abs{F_{xy}({\bf k})}/2$ for wavevector ${\bf k}$ at which the optical transition occurs. We note that this condition holds for a two-band gapped Dirac system at $k=0$:
\begin{align}
    g_{xx}({\bf k}=0)=\abs{F_{xy}({\bf k}=0)}/2. \label{eq:Dirac_MCD_condition}
\end{align}

As illustrated in Fig.~\ref{fig2}(d), the low-energy optical excitation of 3SL MnBi$_2$Te$_4$ only involves two bands, each originating from the Dirac cone surface states. In particular, the transitions occurring around $\hbar\omega\approx 85$ meV is at $\vb{k}=0$ where Eq.~\eqref{eq:Dirac_MCD_condition} is satisfied; hence the $\approx 100\%$ MCD is observed. At higher frequencies, the optical transitions occur away from $\vb{k}=0$, and the $\approx 100\%$ MCD is no longer realized. Nevertheless, due to the unique electronic structure of 3SL MnBi$_2$Te$_4$, the absorption probability of the LCP is still an order of magnitude lower than RCP up to $\hbar\omega \approx 150$ meV. We expect our observation of an enhanced perfect MCD to be valid in thicker MnBi$_2$Te$_4$ films with an odd number of SL (5SL, 7SL, etc., see SM), where the Chern insulator phase has been experimentally observed~\cite{mbt_qah}. To the best of our knowledge, such an enhanced almost perfect MCD in a reasonably wide photon energy window ($65\lesssim \hbar\omega \lesssim 150$ meV) has not been reported so far for a single-phase solid state material. 

\vspace{2cm}
\textbf{High frequency response and the quantum weight}

\begin{figure*}[t]
\includegraphics[width=0.9\textwidth]{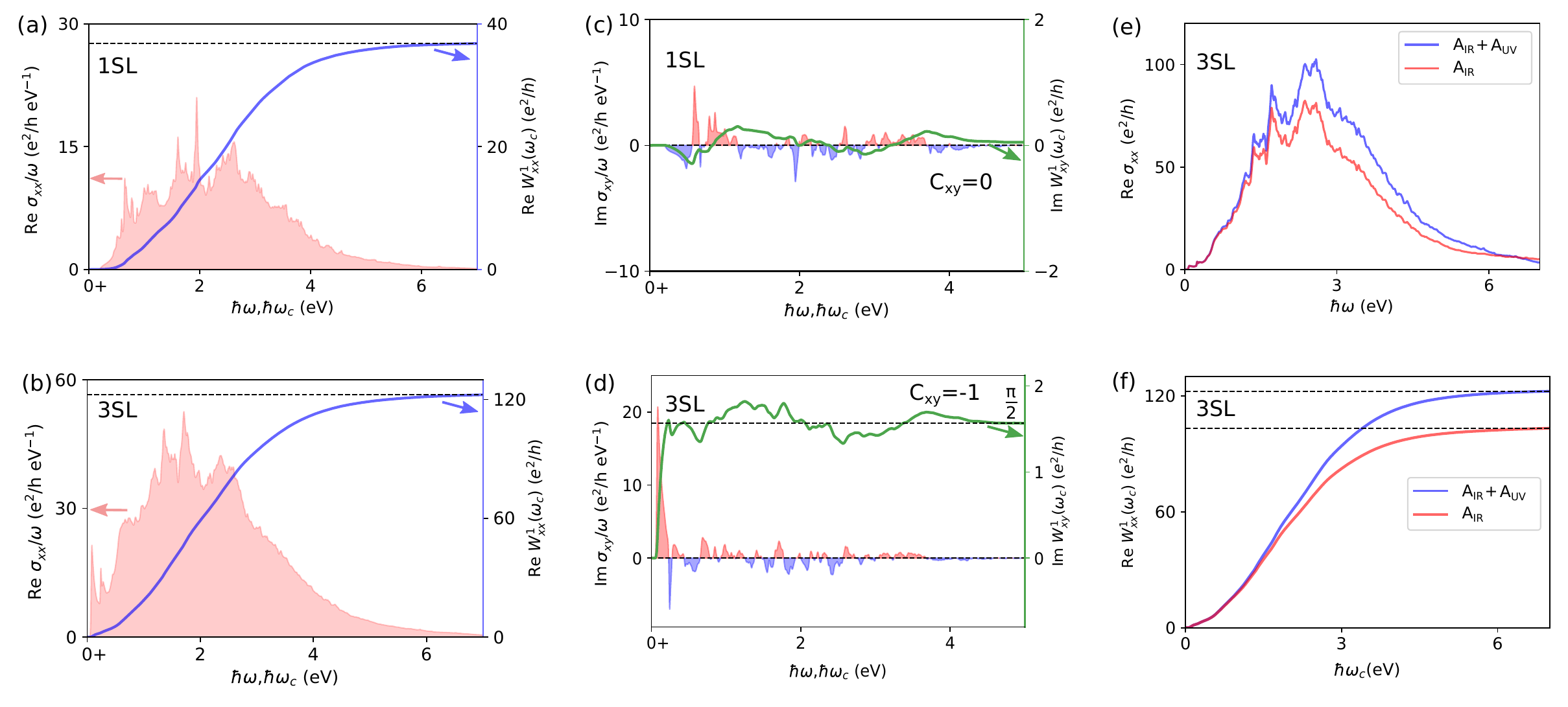}
\caption{{\bf Optical conductivity and optical weight of a few SL thick MnBi$_2$Te$_4$ in a wide photon energy window.} (a,b) The ${\rm Re}~\sigma_{xx}/\omega$ (${\rm Re}~W^1_{xx}$) over a wide photon energy  (cutoff energy) window for (a) 1SL and (b) 3SL MnBi$_2$Te$_4$. The ${\rm Re}~W^1_{xx}$ steadily converges to the quantum weight, the quantum metric of the occupied band manifold integrated over the BZ. (c,d) The ${\rm Im}~\sigma_{xy}/\omega$ $({\rm Im}~W^1_{xy})$ over a wide photon energy (cutoff energy) window for (c) 1SL and (d) 3SL MnBi$_2$Te$_4$. In the case of 1SL, the ${\rm Im}~W^1_{xy}$ converges to zero, while for the 3SL, the ${\rm Im}~W^1_{xy}$ converges to $e^2/4\hbar$, revealing their trivial and the non-trivial ground state, respectively. (e,f) Effect of ignoring the UV part (${\bf A}^{UV}$) of the interband Berry connection on (e) ${\rm Re}~\sigma_{xx}$, and (f) ${\rm Re}~W_{xx}^1$.}
\label{fig6}
\end{figure*}

We now turn the discussion towards the high-frequency optical response of a few SL-thick MnBi$_2$Te$_4$. We study the optical conductivity $\sigma^{abs}_{\alpha\beta}$, and calculate the optical weight $W^1_{\alpha\beta}$ by varying the cutoff frequency in a wide frequency range. First, we focus on the real part of $\sigma^{abs}_{\alpha\beta}$. The high-frequency behavior of ${\rm Re}~\sigma_{xx}$ is presented in the SM. The ${\rm Re}~\sigma_{xx}$ at high-frequency has overall similar characteristics irrespective of the number of layers: it steadily increases with the frequency and reaches its maximum value when the number of available optical transition channels is maximum, before reducing. As shown in Fig.~\ref{fig6}(a-b), for 1SL and 3SL at high frequency, the ${\rm Re}~\sigma_{xx}/\omega$ largely follows the nature of ${\rm Re}~\sigma_{xx}$: it increases steadily, reaches the peak value, and reduces. We evaluate the optical weight ${\rm Re}~W^1_{xx}$ following Eq.~\ref{weight} by varying the cutoff frequency. As evident from Fig~\ref{fig6}(c)-(d), the ${\rm Re}~W^1_{xx}$ steadily increases with the cutoff frequency and converges for $\hbar\omega_c > 6$ eV. From this converged value of ${\rm Re}~W^1_{xx}$, we deduce the quantum weight $K_{xx}$ using Eq.~\ref{EqnsigmaL}. The total quantum weight $K$ is given by $K=2K_{xx}$ due to the $\mathcal{C}_{3z}$ rotation symmetry. For 1SL and 3SL MnBi$_2$Te$_4$, we obtain the total quantum weight $K^{\rm 1SL}=23.42$, and $K^{\rm 3SL}=77.89$, respectively. Although the $\sigma_{xy}$ vanishes identically at all frequencies, the ${\rm Re}~\sigma_{xx}$ leads to a non-zero quantum weight of $K^{\rm 2SL}=50.42$ for 2SL MnBi$_2$Te$_4$ (see SM). 

We now focus on the imaginary part of $\sigma^{abs}_{\alpha\beta}$. The ${\rm Im}~\sigma_{xy}$ has a bounded oscillatory pattern that can take both negative and positive values (see SM). As shown in Fig.~\ref{fig6}(c)-(d), for 1SL, the ${\rm Im}~\sigma_{xy}/\omega$ shows a gradual increase in the low frequency and an oscillatory nature at high frequency. In contrast, for 3SL, the ${\rm Im}~\sigma_{xy}/\omega$ has a large peak at low frequency and shows a diminishing oscillatory pattern at high frequency. The corresponding optical weight, ${\rm Im}~W^1_{xy}$, for 1SL starts from zero and oscillates around the horizontal axis before reaching the final convergence around $\hbar\omega_c\approx5 $ eV. In contrast, in 3SL MnBi$_2$Te$_4$, the ${\rm Im}~W^1_{xy}$ starts from zero, quickly approaches the quantized value of $e^2/4\hbar$, and oscillates around $y=e^2/4\hbar$, revealing the Chern insulator ground state ($C_{xy}=-1$) of 3SL. Evidently, the ${\rm Im}~W^1_{xy}$ can differentiate between the trivial and non-trivial insulating ground state. 

The ${\rm Re}~\sigma_{xx}$ is related to the optical absorption and quantum metric, and it is always positive, resulting in a slower convergence of ${\rm Re}~W^1_{xx}$ with the cutoff frequency. On the other hand, ${\rm Im}~\sigma_{xy}$ is related to the MCD and Berry curvature, and it can be both positive and negative. The Berry curvature effect dominates in the low-frequency region when the optical transitions mostly involve the topological bands that carry large Berry curvature. Consequently, the ${\rm Im}~W^1_{xy}$ converges rapidly within a small cutoff frequency. Clearly, the converged quantum weight far exceeds the lower bound provided by the quantized Chern number in a real material.

It is worth noting that while computing the optical response through Peierls substitution using a tight-binding model, part of the contribution arising from the off-diagonal position matrix elements of the atomic orbitals is generally not included. In a tight-binding model, electrons are assumed to be bound to the atomic sites, and the optical response arises only from the hopping of the electrons. In such a scenario, if the hopping of the tight-binding model is completely turned off, the optical response vanishes. However, in real systems, the optical response can be finite even for an isolated array of atoms without any hopping, where the optical response arises from the transition between different atomic orbitals of the same atom. This contribution becomes important at high frequency. To capture this effect, the total interband Berry connection can be split into two parts: ${\bf A}={\bf A}^{IR}+{\bf A}^{UV}$ (see SM for details). The ${\bf A}^{IR}$ arise due to the hopping of the electrons, while the ${\bf A}^{UV}$ is related to the optical excitation involving orbitals of individual atoms. We highlight the effect of ignoring the ${\bf A}^{UV}$ term on the ${\rm Re}~\sigma_{xx}$ and the resultant quantum weight in Fig.~\ref{fig6}(e-f). It is clear from our results that the ${\bf A}^{UV}$ part has a non-negligible contribution to ${\rm Re}~\sigma_{xx}$ at the high frequency, and therefore it can influence the converged value of the quantum weight significantly. The ${\bf A}^{UV}$ part does not significantly contribute to ${\rm Im}~\sigma_{xy}$ or equivalently to the Chern number (see SM). The non-zero Chern number arises from the low-energy topological bands, and therefore, the ${\bf A}^{IR}$ contribution is sufficient to saturate the quantized Chern number. It should be noted that the ${\bf A}^{UV}$ contribution also depends on the spread of the Wannier function, which can be optimized by performing maximum localization procedures, and this contribution is not unambiguous. Therefore, it is important to include the UV contribution for an unambiguous determination of the quantum weight.

To summarize, using first-principles calculations and effective field theory, we demonstrate how the absorptive part of the optical conductivity in a real material is connected to the ground state quantum geometry and topology. In a quantum anomalous Hall insulator, where the Berry curvature is often highly concentrated around a small region in the momentum space, the optical weight of the MCD within a narrow frequency range can be sufficient to saturate the quantized Chern number. We show that 3SL MnBi$_2$Te$_4$ in the Chern insulator state exhibits enhanced almost perfect MCD in a narrow photon energy range in the infrared region. We quantify the quantum weight for a real material and show that it far exceeds the lower bound provided by the Chern number. Our results connect the optical response to {\it ground-state quantum geometry and topology} in real materials that can motivate designing next-generation optoelectronic devices exploiting the quantum geometric aspect of the topological states of matter. 

\vspace{0.5cm}
\textbf{Author Contributions:} All the authors contributed to the intellectual content of this work. BG initiated the project in discussion with HL and LF. BG performed the first principles calculations and numerical simulations under the supervision of HL, LF, and AB. YO performed the model calculations under the supervision of LF. S-YX shared experimental points of view. BG wrote the original draft with inputs from YO. LF and AB revised the draft. HL, LF, and AB are responsible for the overall direction, planning, and integration among different research units.

\vspace{0.5cm}
\textbf{Acknowledgments:} We thank Junyeong Ahn, and Nabil Atlam for their helpful discussions. The work at Northeastern University was supported by the Air Force Office of Scientific Research under award number FA9550-20-1-0322, and it benefited from the computational resources of Northeastern University's Advanced Scientific Computation Center (ASCC) and the Discovery Cluster. The work at Massachusetts Institute of Technology was supported by the U.S. Army Research Laboratory and the U.S. Army Research Office through the Institute for Soldier Nanotechnologies under Collaborative Agreement Number W911NF-18-2-0048. YO is grateful for the support provided by the Funai Overseas Scholarship. Work in the SYX group was partly supported by the U.S. Department of Energy (DOE) Office of Science through the Ames National Laboratory under contract DE-AC0207CH11358, and partly through Air Force Office of Scientific Research (AFOSR) grant FA9550-23-1-0040. HL acknowledges the support of the National Science and Technology Council (NSTC) in Taiwan under grant number MOST 111-2112-M-001-057-MY3. LF was supported by the Simons Investigator Award from the Simons Foundation. 
\bibliographystyle{Science}
\bibliography{opticalweight.bib}

\end{document}